\documentclass[aps,groupaddress,superscriptaddress,reprint,amsmath,amssymb,graphicx,onecolumn]{revtex4-2}
\usepackage{physics}
\usepackage{bm}
\usepackage{bbold}
\usepackage{graphicx}
\usepackage[toc,page]{appendix}
\usepackage{float}
\usepackage{color}
\usepackage{cleveref}

\newcommand{\avg}[1]{\langle #1 \rangle}

\begin{document}
\preprint{APS/123-QED}

\title{Stochastic processes with multiple temporal scales:\\timescale separation and information}

\author{Giorgio Nicoletti}
\affiliation{Quantitative Life Sciences section, The Abdus Salam International Center for Theoretical Physics (ICTP), Trieste, Italy}
\author{Daniel Maria Busiello}
\affiliation{Max Planck Institute for the Physics of Complex Systems, Dresden, Germany}
\affiliation{Department of Physics and Astronomy ``Galileo Galilei'', University of Padova, Padova, Italy}

\begin{abstract}
\noindent Complex systems are often characterized by the interplay of multiple interconnected dynamical processes operating across a range of temporal scales. This phenomenon is widespread in both biological and artificial scenarios, making it crucial to understand how such multiscale dynamics influence the overall functioning and behavior of these systems. Here, we present a general timescale separation approach that is valid for any set of stochastic differential equations coupled across multiple timescales. We show two alternative derivations, one based on an iterative procedure, and the other grounded on an a priori expansion in relevant small parameters associated with faster timescales. We provide an explicit expression for the conditional structure of the joint probability distribution of the whole set of processes that is solely determined by the multiscale interactions. This result has important consequences in determining how information is generated in the system and how it propagates across different timescales. To demonstrate that our findings are valid independently of the specific dynamical model, we test them against four different (linear and non-linear) dynamics. Then, we focus on the scenario in which two degrees of freedom are reciprocally coupled across their timescales. In this case, we show that the statistics of the fastest dynamics is captured by an effective self-interaction term in the slowest one, creating a regulatory loop. Finally, a connection with analogous previous results obtained for discrete-state systems with higher-order interactions is drawn, elucidating similarities and differences in the physical interpretation of the system. The presented framework might open the avenue for a clearer understanding of the interplay between the underlying multiscale structure and emergent functional behavior of biological and artificial systems.
\end{abstract}

\maketitle

\section{Introduction}
\noindent Understanding biological and nonbiological complex systems often needs the study of processes taking place across multiple spatiotemporal scales, ranging from neuroscience \cite{clark2024theory,timme2014multiplex,cho2016optogenetic,faskowitz2022edges, wolff2022intrinsic} to ecology \cite{hastings2010timescales,bairey2016high,nicoletti2023emergent,sireci2024statistical} and cell signaling \cite{celani2011molecular,ma2009defining,rahi2017oscillatory,tu_chemotaxis}. Indeed, one of the main aspects that makes a system complex - though not the only one - is the presence of several interconnected processes, usually stochastic in nature, that operate at different spatial and temporal levels. Examples are manifold: large biochemical networks composed of modules with their associated timescales \cite{jeong2000large,avanzini2023circuit,liang2022emergent}, signaling networks exhibiting adaptation and habituation \cite{tu_adaptation,nicoletti2023information,kollman_designadapt}, ecological systems exposed to everchanging and noisy environments \cite{hastings2010timescales,villa2019bet}, neural systems characterized by the timescales of neuronal populations, synaptic plasticity, and external signals \cite{clark2024theory,timme2014multiplex,barzon2025excitation,miconi2025neural,mariani2022disentangling}, biological systems transducing information from hidden degrees of freedom \cite{cheong2011information,nicoletti2024tuning}, and even gene-regulatory networks \cite{tkavcik2011information,pham2024dynamical,matsushita2023generic}. A crucial aspect of these systems is often their ability to process information, i.e., to convert certain external or internal inputs into desired outputs through a series of processing operations that are determined by the multiscale structure \cite{bauer2023information,nicoletti2024information,tkavcik2016information}. In recent years, information theory has been proven to provide powerful tools to quantify the performance of such input-output conversion \cite{tostevin2009mutual,nicoletti2021mutual,tostevin2010mutual,nicoletti2022mutual,sachdeva2021optimal,flatt2023abc,amano2022insights}. However, the intimate connection between the multiscale stochastic nature of complex systems and their ability to process information is far from being fully understood, especially when nonlinear dynamics are involved \cite{barzon2025excitation,nicoletti2024integration}.

In this manuscript, we tackle these challenges by introducing a general formalism to study multiscale stochastic systems based on a timescale-separation procedure for an arbitrary number of timescales, an approach we previously developed for discrete systems \cite{nicoletti2024information} and continuous Gaussian processes \cite{nicoletti2024gaussian}. Within this timescale separation framework, we will elucidate the general principles linking interactions between processes at different timescales to the propagation of information throughout the whole system, independently of the underlying dynamics. We quantify this statistical coupling via the Mutual Information Matrix for Multiscale Observables (MIMMO), and test our results with several dynamical systems, both linear and nonlinear. While interactions can be thought of as a structural feature of the system, connections that propagate or generate information play a functional role in both biological and nonbiological systems. Finally, we provide a few examples of systems with pairwise interactions, where the multiscale structure can be described through a multilayer network \cite{dedomenico2013mathematical,bianconi2018multilayer}, highlighting those cases from which nontrivial information-theoretic, functional properties emerge.

\section{Timescale separation for Fokker-Planck operators}
\noindent In this section, we introduce the formalism to describe a multiscale system governed by a set of stochastic differential equations and derive the probability distribution of its degrees of freedom. At the end, we also draw a parallel between this scenario and the one for discrete-state systems with higher-order interactions \cite{grilli2017higher,battiston2021physics,stonge2021universal}.

\subsection{General formalism for Langevin equations with multiple timescales}
\noindent Consider a system composed of $M$ coupled degrees of freedom - e.g., particle positions, neural activities, or chemical concentrations - denoted by $y_m \in \mathbb{R}$. These variables evolve in a generic force field and are subject to a noise coming from the environment, arising for instance from thermal fluctuations or low copy numbers. Their dynamics can be written as the following set of Langevin equations:
\begin{equation}
\label{eqn:langevin_original}
    \dot{y}_m(t) = F_m\left(\bm{y}(t)\right) + \sqrt{2 \tilde{D}_p(\bm{y})} \, \xi_m(t), \quad  m = 1, \dots, M
\end{equation}
where $\bm{y} = (y_1, \dots, y_M)$, $F_m(\bm{y})$ is the $m$-th component of the force field, $\tilde{D}_p(\bm{y})$ is the variable-dependent diffusion coefficient of the $m$-th particle, and $\xi_m$ is a Gaussian white noise that satisfies
\begin{equation}
    \avg{\xi_m(t)} = 0, \quad \avg{\xi_m(t)\xi_{m'}(t')} = \delta_{mm'}\delta(t-t') \;.
\end{equation}
We now assume that the dynamics of the system takes place on $N$ different timescales, which we denote with $\tau_\mu$ for $\mu = 1, \dots, N$. In particular, each timescale characterizes the dynamics of a cluster of $M_\mu$ degrees of freedom, denoted by $\bm{x}_\mu \in \mathbb{R}^{M_\mu}$, with $\sum_{\mu} M_\mu = M$ (Fig.~\ref{fig:multiscale_example}). Without loss of generality, we arbitrarily order the variables so that:
\begin{equation}
    \bm{x}_\mu = \left(y_{i_\mu + 1}, \dots, y_{i_\mu + M_\mu}\right), \quad i_\mu = \sum_{\nu = 1}^{\mu-1}M_\mu \; .
\end{equation}
For the deterministic part of \eqref{eqn:langevin_original}, this amounts to the fact that the force field acting on $\bm{x}_\mu$ can be rescaled as
\begin{equation}
     \bm{F}_\mu\left(\bm{y}\right) = \left( F_{i_\mu + 1} \left(\bm{y}\right), \dots, F_{i_\mu + M_\mu} \left(\bm{y} \right) \right) = \frac{\bm{f}_\mu\left(\bm{x}_1, \dots, \bm{x}_N\right)}{\tau_\mu} \; .
\end{equation}
Similarly, the aggregate variable $\bm{x}_\mu$ follows a Langevin dynamics with a diffusion matrix $\tilde{D}_\mu = \tilde{\sigma}_\mu^T \tilde{\sigma}_\mu \in \mathbb{R}^{M_\mu \times M_\mu}$. Since the diffusive part in \eqref{eqn:langevin_original} scales as $\sim t^{1/2}$, the rescaling takes the following form:
\begin{equation}
\label{eqn:scaling_diffusion}
    \tilde{\sigma}_\mu = \frac{\sigma_\mu}{\sqrt{\tau_\mu}} \qquad \sigma_\mu^T \sigma_\mu = D_\mu \;.
\end{equation}
These choices ensure that the Fokker-Planck operator scales correctly with its associated timescale. As an additional note, we stress that the scaling of the diffusion matrix proposed in Eq.~\eqref{eqn:scaling_diffusion}, from a thermodynamic perspective, ensures that the system is at equilibrium in the presence of reciprocal forces, even if they act on different timescales (for a more in-depth discussion of this aspect, see \cite{nicoletti2024tuning}).

Thus, we can rewrite \eqref{eqn:langevin_original} in terms of the aggregate variables as follows:
\begin{equation}
\label{eqn:langevin_clusters}
    \tau_\mu \dot{\bm{x}}_\mu(t) = \bm{f}_\mu \left(\bm{x}_1(t), \dots, \bm{x}_N(t)\right) + \sqrt{2 \tau_\mu} \sigma_\mu\left(\bm{x}_1(t), \dots, \bm{x}_N(t)\right) \, \bm{\xi}_\mu(t), \qquad \mu = 1, \dots, N
\end{equation}
with $\bm{\xi}_\mu = \left( \xi_{i_\mu+1}, \dots, \xi_{i_\mu + M_\mu} \right)$, or, component-wise,
\begin{equation}
\label{eqn:langevin_clusters_comp}
    \tau_\mu \dot{x}_\mu^i(t) = f_\mu^i \left(\bm{x}_1(t), \dots, \bm{x}_N(t)\right) + \sqrt{2 \tau_\mu D_\mu^i\left(\bm{x}_1(t), \dots, \bm{x}_N(t)\right)} \, \xi_\mu^i(t), \quad i = 1, \dots, M_\mu, \quad \mu = 1, \dots, N \; .
\end{equation}
Here, $D_\mu^i$ denotes the diffusion coefficient of the $i$-th degree of freedom in the $\mu$-th cluster, $f_\mu^i$ the corresponding force field, and we made the timescale manifest on the l.h.s. of the equation. Unless otherwise specified, we will typically use Greek indices to denote the different clusters of particles evolving on the same timescale, and Latin indices to denote the degrees of freedom within a cluster.
Eq.~\eqref{eqn:langevin_clusters} can be equivalently cast as the following Fokker-Planck equation describing the evolution of the joint probability density function of all variables, $p_{1, \dots, N}(\bm{x}_1, \dots, \bm{x}_N, t)$:
\begin{equation}
\label{eqn:FPE}
    \pdv{}{t} p_{1, \dots, N}(\bm{x}_1, \dots, \bm{x}_N, t) = \sum_{\mu = 1}^N \frac{1}{\tau_\mu} \mathcal{L}_\mu(\bm{x}_1, \dots, \bm{x}_N) \, p_{1, \dots, N}(\bm{x}_1, \dots, \bm{x}_N, t) \;,
\end{equation}
with the normalization condition
\begin{equation}
    \int \prod_{\mu = 1}^N \prod_{i = 1}^{M_\mu} d x_\mu^i \,  p_{1, \dots, N}(\bm{x}_1, \dots, \bm{x}_N, t) = 1 \quad \forall t
\end{equation}
and where we introduced the Fokker-Planck operators
\begin{equation}
    \mathcal{L}_\mu(\bm{x}_1, \dots, \bm{x}_N) = - \sum_{i = 1}^{M_\mu} \pdv{}{x_\mu^i} f_\mu^i \left(\bm{x}_1, \dots, \bm{x}_N\right) + \sum_{i = 1}^{M_\mu} \pdv[2]{}{{x_{\mu}^{i}}} D_\mu^i\left(\bm{x}_1, \dots, \bm{x}_N\right) \; .
\end{equation}
In full generality, Eq.~\eqref{eqn:FPE} cannot be solved exactly. However, by exploiting the multiscale structure of the system, and employing a suitable timescale-separation procedure, we will show that it is still possible to uncover several properties of the joint probability $p_{1, \dots, N}$, even without an analytic solution. These features will solely depend on timescale ordering and the interactions between clusters, and are independent of the underlying dynamics.

\subsection{Order-by-order solution with three timescales}
\noindent To illustrate our approach, we start with the simple case of three timescales, i.e., $\tau_1$, $\tau_2$, and $\tau_3$, and three corresponding multidimensional variables, i.e., $\bm{x}_1$, $\bm{x}_2$, and $\bm{x}_3$. The Fokker-Planck equation governing their evolution can be written explicitly as follows:
\begin{equation}
\label{eq:FPE_3}
    \pdv{}{t} p_{1,2,3} = \left( \frac{\mathcal{L}_1}{\tau_1} + \frac{\mathcal{L}_2}{\tau_2} + \frac{\mathcal{L}_3}{\tau_3} \right) p_{1,2,3} \;,
\end{equation}
where, for brevity, we did not write all dependencies in the Fokker-Planck operators. We assume that the dynamical timescales are infinitely different from one another and ordered such that $\tau_1 \ll \tau_2 \ll \tau_3$. Hence, their ratios can be treated as small parameters, i.e.,
\begin{equation}
    \epsilon_\mu = \frac{\tau_\mu}{\tau_3} \;, \quad \mu = 1, 2 \, : \qquad \epsilon_1 \ll \epsilon_2 \ll \epsilon_3 = 1 \;.
\end{equation}
Eq.~\eqref{eq:FPE_3} can be rewritten in terms of these quantities by multiplying on both sides by $\tau_3$ and applying the transformation $t \to \tau = t/\tau_3$. In this way, we are tracking the evolution of the system in the time unit of the slowest timescale. We will show that this choice is compatible with the form of the solution we will obtain. The resulting equation is:
\begin{equation}
    \label{eq:FPE_3_time}
    \pdv{}{\tau} p_{1,2,3} = \left( \frac{\mathcal{L}_1}{\epsilon_1} + \frac{\mathcal{L}_2}{\epsilon_2} + \mathcal{L}_3 \right) p_{1,2,3} \;.
\end{equation}
We now build a timescale-separation ansatz for the joint probability $p_{1,2,3}$ by first expanding on the fastest timescale,
\begin{equation}
\label{eq:p_exp_1}
    p_{1,2,3}(\bm{x}_1, \bm{x}_2, \bm{x}_3, t) = p^{(0)}_{1,2,3}(\bm{x}_1, \bm{x}_2, \bm{x}_3, t) + \epsilon_1 \, p^{(1)}_{1,2,3}(\bm{x}_1, \bm{x}_2, \bm{x}_3, t) + \mathcal{O}(\epsilon_1^2)
\end{equation}
which is a first-order expansion in the smallest parameter $\epsilon_1$ up to the first order \cite{nicoletti2024information,busiello2020coarse,bo2017multiple}. By inserting Eq.~\eqref{eq:p_exp_1} into Eq.~\eqref{eq:FPE_3_time}, and splitting the equations order by order, we obtain the following relationships governing the evolution of the leading contribution to the joint distribution:
\begin{alignat}{3}
\label{eq:FPE_3_order_m1}
&\text{order }\epsilon_1^{-1} & \qquad & 0 = \frac{1}{\epsilon_1}\, \mathcal{L}_1 \, p^{(0)}_{1,2,3} \;, \\
&\text{order }\epsilon_1^{0} & \qquad & \pdv{}{\tau} p^{(0)}_{1,2,3} = \mathcal{L}_1\, p^{(1)}_{1,2,3}
+ \left( \frac{\mathcal{L}_2}{\epsilon_2}
+ \mathcal{L}_3 \right) p^{(0)}_{1,2,3} \;.
\label{eq:FPE_3_order_0}
\end{alignat}
Solving the leading order first, we have that $p^{(0)}_{1,2,3}$ is the stationary solution of the operator $\mathcal{L}_1$, indicating that the fastest degree of freedom is always at steady-state under this approximation. We remark that $\mathcal{L}_1 = \mathcal{L}_1(\bm{x}_1, \bm{x}_2, \bm{x}_3)$, by construction, is a differential operator only for the variable $\bm{x}_1$, while it depends parametrically on $\bm{x}_2$ and $\bm{x}_3$. Thus, if we decompose $p^{(0)}_{1,2,3}$ as 
\begin{equation}
    p^{(0)}_{1,2,3} = p^{(0)}_{1|2,3}\left(\bm{x}_1|\bm{x}_2,\bm{x}_3\right) p^{(0)}_{2,3}\left(\bm{x}_2,\bm{x}_3\right)
    \label{eq:m1_sol}
\end{equation}
we only need to ask that the conditional probability $p^{(0)}_{1|2,3}$ satisfies Eq.~\eqref{eq:FPE_3_order_m1} for every value of $\bm{x}_2$ and $\bm{x}_3$. Therefore, we have that $p^{(0)}_{1|2,3}$ must obey the stationary equation
\begin{equation}
    \mathcal{L}_1 \, p^{\mathrm{st}}_{1|2,3} = 0
\end{equation}
where we dropped the order superscript for brevity, and $(\cdot)^\mathrm{st}$ denotes stationarity with respect to $\mathcal{L}_1$.

Then, to determine $p^{(0)}_{2,3}$, we substitute Eq.~\eqref{eq:m1_sol} into Eq.~\eqref{eq:FPE_3_order_0}:
\begin{equation}
     p^{\mathrm{st}}_{1|2,3} \pdv{}{\tau} p^{(0)}_{2,3} = \mathcal{L}_1\, p^{(1)}_{1,2,3} + \left( \frac{\mathcal{L}_2}{\epsilon_2} +  p^{\mathrm{st}}_{1|2,3} \mathcal{L}_3 \right) p^{(0)}_{2,3} \;.
\end{equation}
Integrating over $\bm{x}_1$, the fastest variable at play, we obtain:
\begin{equation}
    \pdv{}{\tau} p^{(0)}_{2,3} = \left( \frac{\langle \mathcal{L}_2 \rangle_{1|2,3}}{\epsilon_2} + \langle \mathcal{L}_3 
    \rangle_{1|2,3} \right) p^{(0)}_{2,3} \;.
\end{equation}
where we introduced the average operators as $\langle (\cdot) \rangle_{1|2,3} = \int d\bm{x}_1 (\cdot) p^{(0)}_{1|2,3}$. We can apply once again an ansatz of the form
\begin{equation}
    p_{2,3}^{(0)}(\bm{x}_2, \bm{x}_3, t) = p^{(0,0)}_{2,3}(\bm{x}_2, \bm{x}_3, t) + \epsilon_2 \, p^{(0,1)}_{2,3}(\bm{x}_2, \bm{x}_3, t) + \mathcal{O}(\epsilon_2^2) \;,
\end{equation}
where the corrections only refer to this second expansion, and the superscript denotes the order of expansion for each index. Solving again order by order, we have
\begin{equation}
    p^{(0,0)}_{2,3}\left(\bm{x}_2,\bm{x}_3\right) = p^{(0,0),\rm eff}_{2|3}\left(\bm{x}_2|\bm{x}_3\right) p^{(0,0)}_3\left({\bm{x}_3}\right)  \quad \implies \quad \overbrace{\mathcal{L}^{\rm eff}_2}^{\langle \mathcal{L} \rangle_{1|2,3}} p^{\rm st, eff}_{2|3} = 0 \;,
\end{equation}
so that the second fastest degree of freedom is stationary with respect to an effective operator, $\mathcal{L}_2^\mathrm{eff}$, which is the original operator averaged over $\bm{x}_1$. Finally, writing down the evolution at the zeroth order in $\epsilon_2$, and integrating over $\bm{x}_2$, we obtain
\begin{equation}
    \pdv{}{\tau} p^{(0,0),\rm eff}_3 = \overbrace{\langle \mathcal{L}_3 \rangle_{1|2,3; \,2|3}}^{\mathcal{L}_3^{\rm eff}} ~p^{\rm eff}_3
\end{equation}
with $p^{(0,0), \rm eff}_3 = p^{\rm eff}_3$ for brevity. The fact that only this probability evolves over time is compatible with the rescaling we applied to the time at the beginning of this timescale-separation procedure. Notice that the effective operator has to be computed, in principle, by integrating over all the faster degrees of freedom with respect to the one we are studying. Putting everything together, we have that at leading order
\begin{equation}
    p_{1,2,3}^{\rm eff} \left(\bm{x}_1,\bm{x}_2,\bm{x}_3, t\right) \approx p^{\rm st}_{1|2,3}\left(\bm{x}_1|\bm{x}_2,\bm{x}_3\right) p^{\rm st, eff}_{2|3}(\bm{x}_2|\bm{x}_3) p^{\rm eff}_3(\bm{x}_3, t)
\end{equation}
where the superscript $\cdot^{\rm eff}$ on the left-hand side solely reflects the presence of the same superscript in the solution. The conditional structure and the presence of an effective operator both reflect the underlying couplings between clusters evolving over different timescales.

\subsection{General expansion with arbitrary timescales}
\noindent While the case of three timescales can be tackled in detail, the presented procedure can be applied to an arbitrary number of timescales. 
Instead of performing the expansion order by order, we can directly expand $p_{1,\dots,N}$ in powers of all small parameters involved, and solve the resulting equations to determine the form of the leading term, $p^{(0,\dots,0)}_{1,\dots,N}$:
\begin{equation}
\label{eq:p_exp_gen}
    p_{1, \dots, N} = p^{(0,\dots,0)}_{1, \dots, N} + \sum_{\mu = 1}^{N-1} \epsilon_\mu ~p^{(1_\mu)}_{1, \dots, N} + \sum_{\mu=1}^{N-1} \sum_{\nu=1,\nu\neq\mu}^{N-1} \epsilon_\mu \epsilon_\nu ~p^{(1_\mu,1_\nu)}_{1, \dots, N} + \epsilon_1^2 ~p^{(2_\mu)}_{1, \dots, N} + \dots
\end{equation}
with $p^{(m_\mu,n_\nu, \dots)}_{1,\dots,N}$ denoting the contribution at order $m$ in the variable $\mu$, order $n$ in the variable $\nu$, and so on. Inserting this expression into the Fokker-Planck equation, we have
\begin{equation}
\label{eq:FPE_expand}
    \partial_\tau p^{(0,\dots,0)}_{1,\dots,N} + \dots = \sum_{\mu=1}^{N-1} \frac{\mathcal{L}_\mu}{\epsilon_\mu} p^{(0,\dots,0)}_{1,\dots,N} + \sum_{\mu,\nu=1,\nu>\mu}^{N-1} \frac{\epsilon_\nu}{\epsilon_\mu} \mathcal{L_\mu} ~p^{(1_\nu)}_{1,\dots,N} + \mathcal{L}_N ~p^{(0,\dots,0)}_{1,\dots,N} + \dots
\end{equation}
where we only retained the dominant terms. We first notice that the leading order is $\mathcal{O}(\epsilon_1^{-1})$, and we remark that $\mathcal{L}_1$ is a differential operator acting on $\bm{x}_1$ alone. Thus, by decomposing $p^{(0,\dots,0)}_{1, \dots, N}$ into the conditional probability for $\bm{x}_1$ and the joint probability for $\bm{x}_2, \dots, \bm{x}_N$, we obtain the stationary equation
\begin{equation}
    0 = \mathcal{L}_1(\bm{x}_1, \dots, \bm{x}_N) \,  p^{(0,\dots,0)}_{1, \dots, N} \quad \implies \quad p^{(0,\dots,0)}_{1, \dots, N} = p_{1|2, \dots, N}^{\rm st} ~p^{(0,\dots,0)}_{2,\dots,N} 
\end{equation}
where $\mathcal{L}_1(\bm{x}_1, \dots, \bm{x}_N) p_{1|2, \dots, N}^{\rm st} = 0$. 

Since we are interested in determining the leading order of expansion in Eq.~\eqref{eq:p_exp_gen}, $p^{(0,\dots,0)}_{1,\dots,N}$, we can ignore all closed equations resulting from Eq.~\eqref{eq:FPE_expand} that involve higher-order terms in the probability distribution. Hence, the next-to-leading order of interest is $\mathcal{O}(\epsilon_2^{-1})$, which can be written as
\begin{equation}
    0 = \mathcal{L}_2 ~p^{\rm st}_{1|2, \dots, N} ~p^{(0,\dots,0)}_{2,\dots,N} \qquad \xrightarrow{\textrm{integrate over $\bm{x}_1$}} \qquad 0 = \langle \mathcal{L}_2 \rangle_{1|2, \dots, N} ~p^{(0,\dots,0)}_{2,\dots,N} := \mathcal{L}_2^\mathrm{eff} ~p^{(0,\dots,0)}_{2,\dots,N}
\end{equation}
where the effective operator $\mathcal{L}_2^\mathrm{eff} = \langle \mathcal{L}_2 \rangle_{1|2, \dots, N}$ is the operator of the second cluster averaged over the faster dynamics of  $\bm{x}_1$. As a consequence, we find that
\begin{equation}
    p^{(0,\dots,0)}_{2,\dots,N} = p_{2|3, \dots, N}^{\rm st, \rm eff} ~p^{(0,\dots,0)}_{3,\dots,N}, \quad \mathcal{L}_2^\mathrm{eff} p_{2|3, \dots, N}^{\rm st, \rm eff} = 0 \; .
\end{equation}
Repeating the same procedure for all $\epsilon_\mu$, with $\mu=1,\dots,N-1$, to determine all the terms constituting the leading contribution to the probability distribution, we have
\begin{equation}
    p^{(0,\dots,0)}_{1,\dots,N} = p_{1|2, \dots, N}^{\rm st} ~p_{2|3, \dots, N}^{\rm st, eff} \dots p_{N-1|N}^{\rm st, \rm eff} ~p^{(0,\dots,0)}_N
\end{equation}
where the conditional probabilities are defined as the stationary distributions of the effective Fokker-Planck operators
\begin{equation}
\label{eqn:FP_timescale_sep_operators_general}
    \mathcal{L}_\mu^\mathrm{eff}(\bm{x}_\mu, \dots, \bm{x}_N) = \langle \mathcal{L}_\mu \rangle_{1|2, \dots, N; \, \dots; \, \mu-1|\mu, \dots, N} = \int d\bm{x}_1 \dots d\bm{x}_{\mu-1} \prod_{\nu = 1}^{\mu - 1}p^\mathrm{eff, st}_{\nu | \nu +1, \dots, N}(\bm{x}_\nu \, | \, \bm{x}_{\nu + 1}, \dots, N) \, \mathcal{L}_\mu(\bm{x}_1, \dots, \bm{x}_N) \; .
\end{equation}
In particular, the temporal evolution of the probability distribution of the slowest cluster satisfies 
\begin{equation}
    \pdv{}{t} p^{(0,\dots,0)}_N(\bm{x}_N, t) = \mathcal{L}_N^\mathrm{eff}(\bm{x}_N) \, p^{(0,\dots,0)}_N(\bm{x}_N, t)
\end{equation}
as we immediately obtain upon integration over all the fastest degrees of freedom $\bm{x}_1, \dots, \bm{x}_N$. Therefore, if we denote $p^{(0,\dots,0)}_N =  p_N^\mathrm{eff}$ and $p^{(0,\dots,0)}_{1,\dots,N} = p^\mathrm{eff}_{1, \dots, N}$, we obtain the leading-order solution
\begin{equation}
\label{eqn:FP_timescale_sep_sol_general}
    p_{1, \dots, N}^{\rm eff}(\bm{x}_1, \dots, \bm{x}_N, t) = p_N^\mathrm{eff}(\bm{x}_N, t) \prod_{\mu = 1}^{N-1}{p_{\mu | \mu+1, \dots, N}^{\rm eff, st}}(\bm{x}_\mu \, | \, \bm{x}_{\mu + 1}, \dots, \bm{x}_N)
\end{equation}
where we replaced the superscript $\cdot^{(0,\dots,0)}$ on the left-hand side with $\cdot^{\rm eff}$. This expression highlights that the explicit temporal dependence is contained in the slowest cluster, since all the others evolve infinitely fast. Yet, the evolution of the slowest variables reflects into all the other probability distributions through their conditional dependencies.


\section{Probability factorization via propagation paths between timescales}
\subsection{Feedback and direct interaction graph}
\begin{figure}
    \centering
    \includegraphics[width=0.8\textwidth]{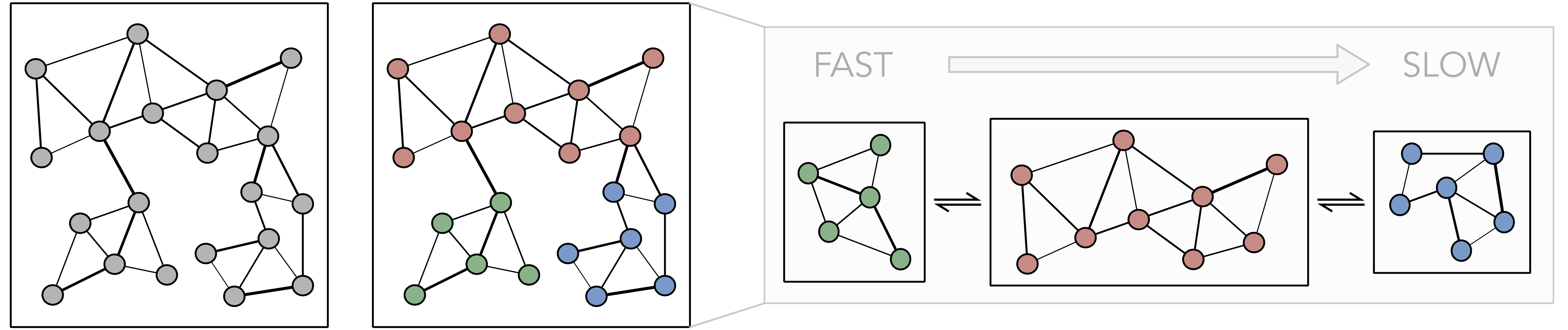}
    \caption{Example of a network with multiple pairwise-connected degrees of freedom cast as a multilayer network in terms of aggregated pairwise-connected variables, where each layer is associated with a given timescale. Connections between layers are pairwise as well by construction.}
    \label{fig:multiscale_example}
\end{figure}

\noindent We now focus on a simpler but general scenario in which interactions between the (aggregated) degrees of freedom $\bm{x}_\mu$ are pairwise. In practice, this can be interpreted as studying a stochastic process defined on the nodes of a directed multilayer network with $N$ layers. Each layer $\mu = 1, \dots, N$ has $M_\mu$ nodes, as before, and is associated with a specific timescale $\tau_\mu$. Layers are connected among themselves by an adjacency matrix $\hat{A}_{\mu\mu}$. We have seen that this setting can be obtaining by coarse-graining a discrete-state system with higher-order interactions. The structure of the multilayer network is described by the adjacency tensor
\begin{equation}
    \hat{A} = \begin{pmatrix}
        \hat{A}_{11} & \hat{A}_{12} & \cdots & \hat{A}_{1N} \\
        \hat{A}_{21} & \hat{A}_{22} & \cdots & \hat{A}_{2N} \\ 
        \vdots & \vdots & \ddots & \vdots \\
        \hat{A}_{N1} & \hat{A}_{N2} & \cdots & \hat{A}_{NN} \\ 
    \end{pmatrix}
\end{equation}
where the elements $A_{\mu\nu}^{ij}$ of the matrix $\hat{A}_{\mu\nu}$ are zero if and only if the $i$-th node of the $\mu$-th layer and the $j$-th node of the $\nu$-th layer are not connected. Otherwise, $A_{\mu\nu}^{ij}$ is the weight of the corresponding edge. We assume that $A_{\mu\mu}^{ii} = 0$. Notice that, for ease of notation in writing dynamical equations, we adopt the index convention that $A_{\mu\nu}^{ij}$ contains the directed connection from the $x_\nu^j$ to $x_\mu^i$ (see Figure \ref{fig:multiscale_example} for a sketch of this scenario).

To fix the ideas, we consider the set of Langevin equations
\begin{align}
\label{eqn:langevin_multilayer}
    \tau_\mu \dot{x}_\mu^i(t) = & - \phi_{\mu}\left(x_\mu^i(t) - \theta_\mu^i\right) + \sum_{j = 1}^{M_\mu} A_{\mu\mu}^{ij}\phi_{\mu\mu}\left(x_\mu^i(t)-\theta_\mu^i, x_\mu^j(t) - \theta_\mu^j\right) + \nonumber \\
    & + \sum_{\nu \ne \mu}\sum_{k = 1}^{M_\nu} A_{\mu\nu}^{ik}\phi_{\mu\nu}\left(x_\mu^i(t)-\theta_\mu^i, x_\nu^k(t)-\theta_\nu^k\right) + \sqrt{2 D_\mu^i \tau_\mu g_\mu(x_\mu)} \, \xi_\mu^i(t) 
\end{align}
where $\theta_\mu^i$ is a bias of the process in the $i$-th node of the $\mu$-th layer, $D_\mu^i$ is its diffusion coefficient, $g_\mu$ is a function modeling the noise amplitude, and $\xi_\mu^i$ are independent white noises. The set of (generic) nonlinear functions $\phi_{\mu\nu}$ describe interactions between the degrees of freedom both within the same timescale and across different timescales. \Cref{eqn:langevin_multilayer} can be described by the Fokker-Planck operators
\begin{equation}
\label{eqn:layer_FP_operators}
    \mathcal{L}_\mu\left(\bm{x}_\mu, \{\bm{x}\}_{\rightsquigarrow \mu}\right) = - \sum_{i = 1}^{M_\mu} \pdv{}{x_\mu^i} \left[\sum_{j=1}^{M_\mu}W_{\mu\mu}^{ij} \phi_{\mu\mu}(x_\mu^j - \theta_\mu^j) - \sum_{\nu \ne \mu}\sum_{k = 1}^{M_\nu} A_{\mu\nu}^{ik}\phi_{\mu\nu}\left(x_\nu^k-\theta_\nu^k\right) \right] + \sum_{i = 1}^{M_\mu} D_\mu^i\pdv[2]{}{{x_{\mu}^{i}}}
\end{equation}
where $\{\bm{x}\}_{\rightsquigarrow \mu}$ is a shorthand notation for the set of all variables $\bm{x}_\nu$ connected to $\mu$, i.e., $\{\bm{x}_\nu\}_{\nu\rightsquigarrow \mu}$ with $\hat{A}_{\mu\nu} \ne 0$, and
\begin{equation}
\label{eqn:Wtensor}
    \hat{W} = \begin{pmatrix}
        \hat{A}_{11} - \mathbb{I}_{1} & \hat{A}_{12} & \cdots & \hat{A}_{1N} \\
        \hat{A}_{21} & \hat{A}_{22} - \mathbb{I}_2 & \cdots & \hat{A}_{2N} \\ 
        \vdots & \vdots & \ddots & \vdots \\
        \hat{A}_{N1} & \hat{A}_{N2} & \cdots & \hat{A}_{NN} - \mathbb{I}_N \\ 
    \end{pmatrix}
\end{equation}
with $\mathbb{I}_\mu$ the $M_\mu \times M_\mu$ identity matrix.

Because of the intrinsic difference between slow and fast variables, here and in the following, it is useful to distinguish interactions between different temporal scales in terms of their directionality with respect to the timescale ordering. Consider the directed graph $\mathcal{G}$ with $N$ nodes associated with the multilayer network, where each node now represents a layer, and edges are given by interactions between layers. In this coarse-grained graph, we name interactions going from a fast layer to a slow layer ``direct interactions'', i.e., edges contained in $\hat{A}_{\mu\nu}$ with $\mu > \nu$. On the other hand, interactions going from a slow to a fast layer are ``feedback interactions'', with $\mu < \nu$. As we will show, these two types of interactions play a different role in determining the conditional dependencies of the solution, further specializing the general form of the probability distribution at the leading order derived in the previous section in full generality.

\begin{figure}
    \centering
    \includegraphics[width=0.8\columnwidth]{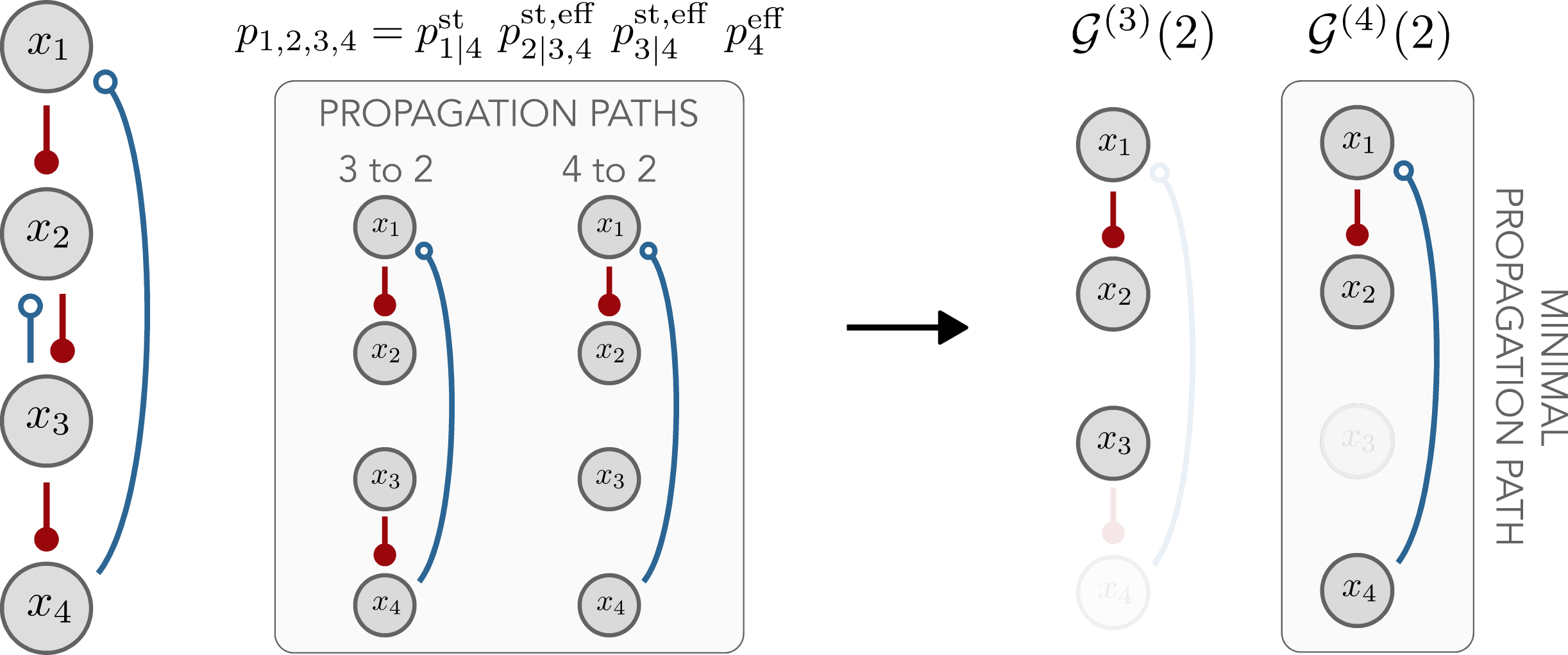}
    \caption{Construction of a minimal propagation path (mPP) for a 4-layer system. We start from the coarse-grained graph $\mathcal{G}$. We are interested in determining $\rho(2)$. We first build all propagation paths (PPs) to layer $2$. Then, we construct the two possible induced subgraphs for this case, $\mathcal{G}^{(3)}(2)$ and $\mathcal{G}^{(4)}(2)$, and show that the one starting from layer $4$ survives and, as such, it is a mPP. Thus, $\rho(2) = \{3, 4\}$, since $4$ belongs to a mPP and $3$ is directly connected to $2$ through a feedback link (blue).}
    \label{fig:mPP}
\end{figure}

\subsection{Paths between timescales and minimal propagation paths}
\noindent We can now retrace the steps to obtain the timescale separation solution in this specific setting. Once again, we assume that the layers are ordered by their timescales, i.e., we take $\tau_1 \ll \tau_2 \dots \ll \tau_N$. As before, the first order is associated with the layer with the fastest dynamics - the first layer, in our ordering - and is immediately solved by the stationary solution of
\begin{equation}
    0 = \mathcal{L}_1\left(\{\bm{x}\}_{\rightsquigarrow 1}\right)\,p^\mathrm{st}_{1 | \rho(1)}(\bm{x}_1 | \{\bm{x}_\nu\}_{\nu \in \rho(1)}) \;.
\end{equation}
We remark on a few main differences in the notation. First, $\mathcal{L}_1$ depends only on a subset of degrees of freedom now, indicated by $\{x\}_{\rightsquigarrow 1}$. The set $\rho(1) \equiv \{x\}_{\rightsquigarrow 1} \in S(1)$ in this case, since only a fraction of all slower degrees of freedom are connected to a node in the layer $1$. For all the layers contained in $\rho(1)$, therefore, there is at least one node connected to a node of the first layer, i.e., in terms of the adjacency tensor,
\begin{equation}
    \rho(1) = \{\nu : \exists \,i,j \, |\, A_{1\nu}^{ij} \ne 0\} \; .
\end{equation}
As before, all these variables enter as conditional dependencies, which amounts to solving the first layer's Fokker-Planck equation while keeping the value of all nodes in all other layers fixed - i.e., by freezing their dynamics. The second order, then, is the stationary solution of the effective operator
\begin{equation}
    \mathcal{L}^\mathrm{eff}_{2} = \int d\bm{x}_1 \mathcal{L}_2\left(\{x\}_{\rightsquigarrow 2}\right) \, p^\mathrm{st}_{1 | \rho(1)}(\bm{x}_1 | \{\bm{x}_\nu\}_{\nu \in \rho(1)})
\end{equation}
explicating the average in this new notation. Thus, we have:
\begin{equation}
    0 = \mathcal{L}^\mathrm{eff}_{2} p^{\mathrm{eff, st}}_{2 | \rho(2)} (\bm{x}_2 | \{\bm{x}_\nu\}_{\nu \in \rho(2)}) \;.
\end{equation}
Here, again $\rho(2) \in S(2)$, but $\rho(2) \neq \{\bm{x}\}_{\rightsquigarrow 2}$. Indeed, since this set of layers must denote all conditional dependencies that remain in the effective Fokker-Planck operator $\mathcal{L}^\mathrm{eff}_{2}$ after the integration over $\bm{x}_1$, it must contain once more all slower layers connected to the second:
\begin{equation}
     \{\nu > 2 : \exists \,i,j \, |\, A_{2\nu}^{ij} \ne 0\} \subset \rho(2).
\end{equation}
However, these are not the only conditional dependencies left. Indeed, $\rho(2)$ may also inherit dependencies on slower layers from all faster layers that have been integrated out - in this case, from $\rho(1)$ appearing in $p^\mathrm{st}_{1 | \rho(1)}$. Therefore, if the first layer is connected to the second, the conditional dependencies contained in $\rho(2)$ will also include all layers directly connected to the first and slower than the second:
\begin{equation}
     \rho(2) = 
     \begin{cases}
         \rho(1) \, \cup \, \{\nu : \exists \,i,j \, |\, A_{2\nu}^{ij} \ne 0\} & \mathrm{if} \quad \{\exists \,i,j \, |\, A_{21}^{ij} \ne 0\} \\
         \{\nu : \exists \,i,j \, |\, A_{2\nu}^{ij} \ne 0\} & \mathrm{otherwise} \\
     \end{cases} \; .
\end{equation}
That is, without interactions from the first to the second layer, the effective operator will not inherit dependencies from $\rho(1)$ and only keep the dependencies coming from slower layers connected to the layer $2$. By iterating this procedure, we can write the hierarchy of effective Fokker-Planck operators as
\begin{equation}
\label{eqn:effective_operators}
    \mathcal{L}^\mathrm{eff}_{\mu} = \int d\bm{x}_1 \dots d \bm{x}_{\mu -1} \mathcal{L}_\mu\left(\{x\}_{\rightsquigarrow \mu}\right) \prod_{\nu = 1}^{\mu - 1} p_{\nu | \rho(\nu)}^{\rm eff, st}
\end{equation}
where $p_{\nu | \rho(\nu)}^{\rm eff, st} (\bm{x}_\nu | \{\bm{x}_\alpha\}_{\alpha \in \rho(\nu)})$ is the stationary probability of the $\nu$-th layer, obeying
\begin{equation}
    \mathcal{L}^\mathrm{eff}_{\nu} \, p_{\nu | \rho(\nu)}^{\rm eff, st} (\bm{x}_\nu | \{\bm{x}_\alpha\}_{\alpha \in \rho(\nu)}) = 0 \; .
\end{equation}
Notice that, for linear interactions, all $p_{\nu | \rho(\nu)}^{\rm eff, st}$ remain Gaussian by definition \cite{nicoletti2024gaussian}. In the previous section, $\rho(\mu) = S(\mu)$, for $\mu = 1, \dots, N$, since we considered the case in which all layers are connected to each other. From now on, we will only refer to $\rho(\mu)$ as this is the most general definition for all layers appearing as conditional dependencies.

By explicitly carrying out the recursive solution, it is possible to interpret the set $\rho(\mu)$ in terms of the topology of the multilayer network. Consider the coarse-grained graph $\mathcal{G}$ and the definition of direct and feedback interactions introduced above. We also define a propagation path (PP) on $\mathcal{G}$ as a directed path from layer $\nu$ to layer $\mu$ with $\nu > \mu$ that passes at most once on each layer, and such that it contains at least one direct interaction, i.e., one edge $\alpha \to \beta$ with $\alpha < \beta$. Then, we consider the graph $\mathcal{G}^{(\nu^*)}(\nu)$ as the induced subgraph obtained by removing all layers slower than $\nu^*$ - i.e., all nodes $\alpha > \nu^*$ - except $\nu$. A propagation path from $\nu$ to $\mu$ is a minimal propagation path if it is a PP both in $\mathcal{G}$ and in the induced subgraph $\mathcal{G}^{(\mu)}(\nu)$. That is, an mPP from $\nu$ to $\mu$ must be a PP on the induced subgraph obtained by removing all layers slower than $\mu$ except $\nu$, strongly constraining the topology (see Fig.~\ref{fig:mPP} for a graphical construction of an mPP). From its definition, it is possible to see that $\rho(\mu)$ is the set
\begin{equation}
\label{eqn:rho}
    \rho(\mu) = \left\{ \nu >\mu : \exists \text{ mPP } \nu \to \mu \text{ or } \hat{A}_{\mu\nu} \ne 0\right\}
\end{equation}
which contains all layers connected to $\mu$ either via an mPP or a single feedback interaction. Thus, by inspecting interactions between layers in the coarse-grained graph $\mathcal{G}$, we can build $\rho(\mu)$ only from the topology of edges between layers, i.e., from the underlying causal structure of the system. Then, the multilayer probability in the timescale separation regime, at the leading order, can be written as
\begin{equation}
\label{eqn:multilayer_probability}
    p_{1, \dots, N}^{\rm eff}(\bm{x}_1, \dots, \bm{x}_N, t) = p_N^\mathrm{eff}(\bm{x}_N, t) \left( \prod_{\nu = 2}^{N-1}p_{\nu | \rho(\nu)}^{\rm eff, st}(\bm{x}_\nu | \{\bm{x}_\alpha\}_{\alpha \in \rho(\nu)}) \right) p_{1 | \rho(1)}^{\rm st}(\bm{x}_1 | \{\bm{x}_\alpha\}_{\alpha \in \rho(1)})
\end{equation}
since the dynamics of the fastest layer is never governed by an effective operator. We notice that the complexity of the conditional dependencies can be cast in terms of minimal propagation paths and feedback interactions that exist across different timescales, i.e., on $\mathcal{G}$.

\section{Mutual information between processes at different timescales}
\subsection{Introducing MIMMO: a Mutual Information Matrix for Multiscale Observables}
\noindent To study how the interactions between different timescales are reflected in the statistical dependencies between the layers, we introduce the mutual information
\begin{equation}
    \label{eqn:mutual_ts}
    I_{\mu\nu}(t) = \int d \bm{x}_\mu d\bm{x}_\nu \, p_{\mu\nu}^\mathrm{eff}(\bm{x}_\mu, \bm{x}_\nu, t) \log_2 \frac{p_{\mu\nu}^\mathrm{eff}(\bm{x}_\mu, \bm{x}_\nu, t)}{p_{\mu}^\mathrm{eff}(\bm{x}_\nu, t)p_{\nu}^\mathrm{eff}(\bm{x}_\nu, t)}
\end{equation}
which captures both linear and nonlinear relationships between the involved variables. Here, the marginal two-point distributions are simply given by
\begin{equation}
    p_{\mu\nu}^\mathrm{eff}(\bm{x}_\mu, \bm{x}_\nu, t) = \int \prod_{\alpha \ne \mu, \nu}^N d \bm{x}_\alpha \, p_N^\mathrm{eff}(\bm{x}_N, t) \prod_{\beta = 1}^{N-1}{p_{\beta | \beta+1, \dots, N}^{\rm eff, st}}(\bm{x}_\beta \, | \, \bm{x}_{\beta + 1}, \dots, \bm{x}_N)
\end{equation}
and similarly for $p_\mu^\mathrm{eff}(\bm{x}_\mu, t)$ and $p_\nu^\mathrm{eff}(\bm{x}_\nu, t)$. Notice that, in full generality, the temporal evolution of all degrees of freedom is inherited from the slowest layer through the conditional dependencies.

\Cref{eqn:mutual_ts} can be computed for every pair of layers, hence defining a $N \times N$ symmetric matrix with positive entries that we name Mutual Information Matrix for Multiscale Observables (MIMMO) \cite{nicoletti2024information}. If, at any time $t$, $I_{\mu\nu} > 0$, the two layers evolving on different timescales are statistically dependent. This implies that the two share information during their temporal evolution and, as such, are coupled together in a \emph{functional} way. Notably, two layers may be strongly interacting but statistically independent due to the effect of their different timescales. In this particular case, we would find $I_{\mu\nu} = 0$. This means that the presence of a coupling does not necessarily reflects into a non-zero mutual information. On the other hand, layers that do not interact directly may become strongly functionally coupled, implying that $I_{\mu\nu} > 0$. In other words, the interplay between the underlying structure of interactions and the timescale ordering determines the statistical dependencies in a stochastic multiscale system. MIMMO provides a tool to capture this effect. 

\subsection{Systems with three timescales}
\noindent We now explore in detail the prime role of the directionality of interactions with respect to the corresponding timescales by studying a three-layer system with four different minimal motifs in $\mathcal{G}$, i.e., of interactions between layers: when only direct interactions are present; when only feedback interactions are present; with a propagation path; and with a minimal propagation path. As we will see, this will allow us to outline some general principles governing how information is generated and propagated across different layers.

We also consider four different types of network dynamics, both linear and nonlinear \cite{harush2017dynamic, meena2023emergent, barzon2024unraveling}:
\begin{align}
    \text{linear dynamics: } \quad & \phi_{\mu}(x_\mu) = x_\mu, \quad \phi_{\mu\nu}(x_\mu, x_\nu) = x_\nu, \quad g_\mu(x_\mu) = 1 \\
    \text{neural dynamics: } \quad & \phi_{\mu}(x_\mu) = x_\mu - \tanh(x_\mu), \quad \phi_{\mu\nu}(x_\mu, x_\nu) = \tanh(x_\nu), \quad g_\mu(x_\mu) = 1 \nonumber\\
    \text{ecological dynamics: } \quad & \phi_{\mu}(x_\mu) = x_\mu(1-x_\mu) + h , \quad \phi_{\mu\nu}(x_\mu, x_\nu) = \Lambda x_\mu x_\nu, \quad g_\mu(x_\mu) = x_\mu \nonumber \\
    \text{contact process: } \quad & \phi_{\mu}(x_\mu) = x_\mu + h, \quad \phi_{\mu\nu}(x_\mu, x_\nu) = \Lambda (1 - x_\mu) x_\nu, \quad g_\mu(x_\mu) = x_\mu(1-x_\mu), \nonumber
    \label{eqn:dynamics_type}
\end{align}
and, for ease of visualization, we plot our results in the case of one-dimensional layers. Notice that, for ecological models, $h$ represents a constant migration rate; for the contact process, $h$ is an external field and $\Lambda$ quantifies the interaction strength.

\subsubsection{Direct interactions path}
\begin{figure}
    \centering
    \includegraphics[width=\textwidth]{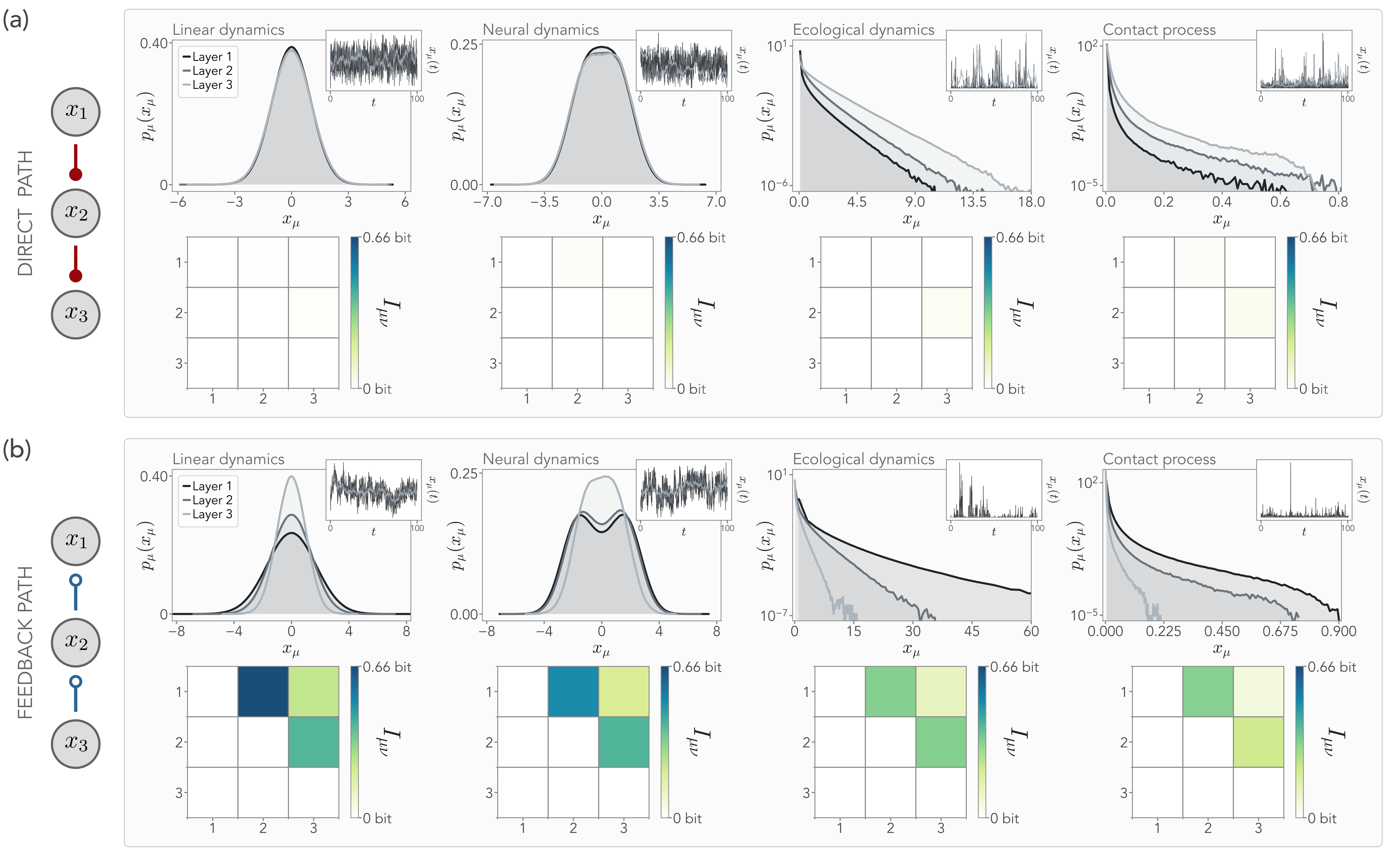}
    \caption{(a) Stationary distributions and MIMMO for a direct interaction path, in which a fast layer influences the next one, for different types of dynamics. In this case, the joint probability between the layers is exactly factorized in the timescale separation limit. Thus, no mutual information between them is present regardless of the dynamics. (b) Same, but for a feedback path, in which a slow layer influences the one before. In this case, the joint probability cannot be factorized, and all layers depend on one another. Thus, all elements of the MIMMO are different from zero. Importantly, different dynamics lead to different values of the mutual information. In this figure, the mutual information was estimated from Langevin trajectories where $\tau_3 = 10$, $\tau_2 = 1$, and $\tau_1 = 0.1$, so that the separation between the timescales is only one order of magnitude, and all diffusion coefficients have been set to $1$. For both ecological and the contact process dynamics, we set $h = 10^{-3}$. For the ecological dynamics, $\Lambda = 2.5$, whereas $\Lambda = 2$ for the contact process one. For ease of visualization, we study a system where all layers are one-dimensional.}
    \label{fig:DD_FF}
\end{figure}

\noindent We begin with a simple multilayer network where the first layer influences the second, and the second influences the third (Figure \ref{fig:DD_FF}a). In this case, all interactions are direct. Hence,
\begin{equation}
    \hat{A} = \begin{pmatrix}
        \hat{A}_{11} & 0 & 0 \\
        \hat{A}_{21} & \hat{A}_{22} & 0 \\
        0 & \hat{A}_{32}  & \hat{A}_{33} 
    \end{pmatrix}
\end{equation}
so that it does not exist any block such that $\hat{A}_{\mu\nu} \ne 0$ for $\mu < \nu$. Similarly, there is no directed path from a slow to a fast layer, i.e., the graph $\mathcal{G}$ has no propagation paths. Thus, by following \cref{eqn:rho}, $\rho(\mu)$ is the empty set for all $\mu$. Then, the solution in the timescale separation regime simply reads
\begin{equation}
    p_{1, 2, 3}^{\rm eff}(\bm{x}_1, \bm{x}_2, \bm{x}_3, t) = p_1^\mathrm{st}(\bm{x}_1) \, p_2^\mathrm{eff, st}(\bm{x}_2) \, p_3^\mathrm{eff}(\bm{x}_3, t)
\end{equation}
from \cref{eqn:multilayer_probability}. Clearly, this solution is already factorized in form, so all elements of the MIMMO are zero:
\begin{equation}
    I_{\mu\nu} = 0 \quad \quad \forall \mu \ne \nu
\end{equation}
which shows that information cannot be generated through direct interactions, i.e., from fast to slow variables. Thus, despite the presence of non-zero interactions, the timescale separation between the layers makes them statistically independent. In Fig.~\cref{fig:DD_FF}a, we show that this is indeed true regardless of the type of dynamics at hand, as expected. Crucially, our results remain qualitatively valid even with a relatively small difference between the timescales, i.e., $\tau_1/\tau_2 = \tau_2/\tau_3 = 10^{-1}$.

\subsubsection{Feedback interactions path}
\noindent We now consider the opposite case, where all layers are connected by feedback interactions only (Figure \ref{fig:DD_FF}b):
\begin{equation}
    \hat{A} = \begin{pmatrix}
        \hat{A}_{11} & \hat{A}_{12} & 0 \\
        0 & \hat{A}_{22} & \hat{A}_{23} \\
        0 & 0  & \hat{A}_{33} 
    \end{pmatrix}
\end{equation}
and, since no direct interactions are present, there is no PP in $\mathcal{G}$ once more. However, from \cref{eqn:rho}, we have
\begin{equation}
    \rho(1) = \{2\}, \quad \rho(2) = \{3\}, \quad \rho(3) = \{\} \; .
\end{equation}
Thus, \cref{eqn:multilayer_probability} tells us that 
\begin{equation}
    p_{1, 2, 3}(\bm{x}_1, \bm{x}_2, \bm{x}_3, t) = p_{1|2}^\mathrm{st}(\bm{x}_1 \, | \, \bm{x}_2) \, p_{2|3}^\mathrm{st}(\bm{x}_2 \, | \, \bm{x}_3) \, p_3(\bm{x}_3, t)
\end{equation}
where no ``effective'' superscript appears as all probabilities in the factorization evolve with their own Fokker-Planck operators, with the other degrees of freedom being fixed according to the timescale separation procedure. Clearly, in this case, we have that
\begin{equation}
    I_{\mu\nu} \ge 0 \quad \quad \forall \mu \neq \nu
\end{equation}
where it is worth noting that the information between $\bm{x}_1$ and $\bm{x}_3$ is relayed by $\bm{x}_2$ through the feedback interactions. The value of the mutual information itself will depend on the specific dynamics at hand - i.e., on the shape of the probability distributions, as we show in Figure \ref{fig:DD_FF}b, but the elements of MIMMO are always non-zero. 

\subsubsection{Propagation path}
\begin{figure}
    \centering
    \includegraphics[width=\textwidth]{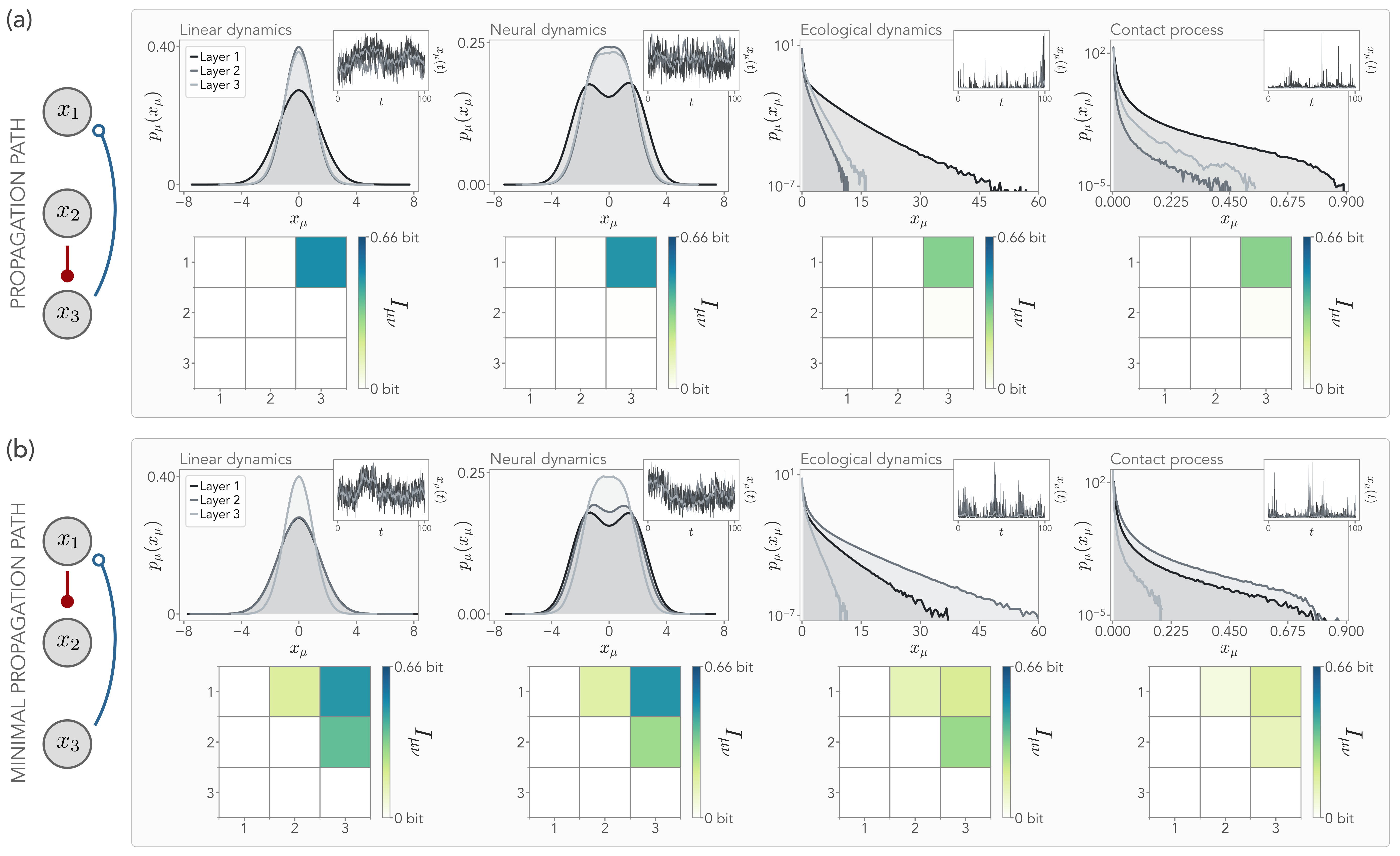}
    \caption{(a) Stationary distributions and MIMMO for a non-minimal propagation path for different types of dynamics. In this case, our solution shows that only the first and the third layer depend on one another in the timescale separation limit. Thus, for all types of dynamics, only the $I_{13}$ element of the MIMMO can be different from zero. (b) Same, but for a minimal propagation path. As in the feedback case, the joint probability cannot be factorized, and all elements of the MIMMO are different from zero. Importantly, different dynamics lead to different values of the mutual information while retaining the predicted structure of the MIMMO. In this figure, the mutual information was estimated from Langevin trajectories where $\tau_3 = 10$, $\tau_2 = 1$, and $\tau_1 = 0.1$, so that the separation between the timescales is only one order of magnitude, and all diffusion coefficients have been set to $1$. For both ecological and the contact process dynamics, we set $h = 10^{-3}$. For the ecological dynamics, $\Lambda = 2.5$, whereas $\Lambda = 2$ for the contact process one. For ease of visualization, we study a system where all layers are one-dimensional.}
    \label{fig:PP_mPP}
\end{figure}

\noindent A propagation path in $\mathcal{G}$ can be found with the structure of interactions shown  in Fig.~\ref{fig:PP_mPP}a, i.e.,
\begin{equation}
    \hat{A} = \begin{pmatrix}
        \hat{A}_{11} & 0 & \hat{A}_{13} \\
        0 & \hat{A}_{22} & 0 \\
        0 & \hat{A}_{32} & \hat{A}_{33} 
    \end{pmatrix}
\end{equation}
so that $\hat{A}_{13}$ describes a feedback interaction from $3$ to $1$, whereas $\hat{A}_{32}$ describes a direct interaction from $2$ to $3$. Therefore, $\bm{x}_2 \to \bm{x}_3 \to \bm{x}_1$ is a PP from $2$ to $1$, but not a mPP.
Indeed, when considering the induced subgraph $\mathcal{G}^{(2)}(1)$, where all layers slower than $1$ have been removed, except for $2$, this is not a PP anymore. Thus, we have
\begin{equation}
    \rho(1) = \{3\}, \quad \rho(2) = \{\}, \quad \rho(3) = \{\}
\end{equation}
so that, following \cref{eqn:multilayer_probability}, we have:
\begin{equation}
    p_{1, 2, 3}^{\rm eff}(\bm{x}_1, \bm{x}_2, \bm{x}_3, t) = p_{1|3}^\mathrm{st}(\bm{x}_1 \, | \, \bm{x}_2) \, p_{2}^\mathrm{st}(\bm{x}_2) \, p_3^\mathrm{eff}(\bm{x}_3, t)
\end{equation}
Therefore, in the MIMMO we have that only $I_{13} \neq 0$, independently of the dynamics considered (see Fig.~\ref{fig:mPP}a). This is intuitively happening because the effect of the direct interaction from $2$ to $3$ is purely dynamical and does not affect information - i.e., the statistical dependence is lost due to the difference in the timescales of the two layers.

\subsubsection{Minimal propagation path}
\noindent Finally, we now study the case of a minimal propagation path (see \Cref{fig:PP_mPP}b). The adjacency tensor is given by
\begin{equation}
    \hat{A} = \begin{pmatrix}
        \hat{A}_{11} & 0 & \hat{A}_{13} \\
        \hat{A}_{21} & \hat{A}_{22} & 0 \\
        0 & 0 & \hat{A}_{33}
    \end{pmatrix} \;,
\end{equation}
so that $\hat{A}_{13}$ describes a feedback interaction from $3$ to $1$, as before, but now $\hat{A}_{21}$ is a direct interaction from $1$ to $2$. In this example, $\bm{x}_3 \to \bm{x}_1 \to \bm{x}_2$ is a PP from $3$ to $2$.
However, when considering the induced subgraph $\mathcal{G}^{(3)}(2)$, it remains a PP and, as a consequence, it is a mPP from $3$ to $1$. Thus, we have:
\begin{equation}
    \rho(1) = \{3\}, \quad \rho(2) = \{3\}, \quad \rho(3) = \{\}
\end{equation}
where the crucial difference in $\rho(2)$ with respect to the previous case leads to a markedly different probability structure:
\begin{equation}
    p_{1, 2, 3}^{\rm eff}(\bm{x}_1, \bm{x}_2, \bm{x}_3, t) = p_{1|3}^\mathrm{st}(\bm{x}_1 \, | \, \bm{x}_3) \, p_{2|3}^\mathrm{eff,st}(\bm{x}_2 \, | \, \bm{x}_3) \, p_3(\bm{x}_3, t) \;.
\end{equation}
We remark that the dependence of $2$ on $3$ in the effective operator $\mathcal{L}_2^\mathrm{eff}$ is inherited through the marginalization over the stationary distribution of $1$, due to the direct interaction from $1$ to $2$, which in turn is conditionally dependent on $3$ due to the feedback interaction with the slow layer. This is precisely the effect of the minimal propagation path - information is generated from a slow layer ($3$) in a fast one ($1$) through a feedback interaction, and is then propagated by direct interactions to layers that are slower than the intermediate one ($1$ in this case) but faster than the initial one ($3$ in this case). As a consequence,
\begin{equation}
    I_{12} \ge 0, \quad I_{13} \ge 0, \quad I_{23} \ge 0 \;.
\end{equation}
Notably, the origin of information between layers is very different since it might stem from feedback links or from the propagation through direct connections.
In \Cref{fig:PP_mPP}b, we show once more that, while the specific values of the MIMMO depend on the dynamics, its structure remains unchanged.

\subsubsection{Information propagation across timescales}
\noindent The previous examples show that the topological structure of interactions between different timescales fully determines the propagation of information. Our results depend only on the general structure of the probability in \cref{eqn:multilayer_probability}, not on the dynamics of the layer nor the form of the interactions, which rather determine how large the MIMMO elements may be. As a consequence, the main outcome of the presented timescale separation approach resides in the identification of three basic physical principles governing the propagation of information across timescales:
\begin{enumerate}
    \item[(i)] feedback interactions generate information from slow to fast layers; 
    \item[(ii)] direct interactions alone do not generate information; 
    \item[(iii)] information generated through feedback by a slow layer may be propagated through direct interactions to any layer faster than the generating one. 
\end{enumerate}
The first two principles elucidate the different roles of feedback and direct interactions in determining the information content of a stochastic multiscale system. In other words, direct interactions do not create statistical dependencies and act only on a dynamical level, whereas feedback interactions lead to nonzero mutual information between the connected layers. The third principle allows us to identify an emergent directionality in information propagation, despite the symmetry of MIMMO. At the same time, it poses serious constraints on the efficacy of control mechanisms in stochastic networks, revealing that they create functional couplings, i.e., mutual information, only when acting as regulatory processes, i.e., from slow to fast layers.

\subsection{Loop between two timescales}
\noindent We now consider the specific case of a feedback loop between two timescales, described by the adjacency tensor 
\begin{equation}
    \hat{A} = \begin{pmatrix}
        \hat{A}_{11} & \hat{A}_{12} \\
        \hat{A}_{21} & \hat{A}_{22}
    \end{pmatrix}
    \label{eqn:adj_loop}
\end{equation}
so that the second layer interacts with the first one via feedback interactions $\hat{A}_{12}$, which in turn influence it back through direct interactions $\hat{A}_{21}$. The two-layer stationary probability is
\begin{equation}
    p_{12}^\mathrm{st}(\bm{x}_1, \bm{x}_2) = p^\mathrm{st}_{1 | 2}(\bm{x}_1 \, | \, \bm{x}_2) p^\mathrm{eff, st}_{2}(\bm{x}_2)
\end{equation}
where the slow layer is governed by the effective Fokker-Planck operator
\begin{equation}
    \mathcal{L}_2^\mathrm{eff}(\bm{x}_2) = \int d\bm{x}_2 p^\mathrm{st}_{1 | 2}(\bm{x}_1 \, | \, \bm{x}_2) \, \mathcal{L}_2(\bm{x}_1, \bm{x}_2) \; .
\end{equation}
In this scenario, the marginalization over the fast layer $\bm{x}_1$ does not introduce additional dependencies in $\mathcal{L}_2^\mathrm{eff}$, but rather acts as a self-interaction for $\bm{x}_2$. Explicitly, 
\begin{equation}
    \mathcal{L}_2^\mathrm{eff}(\bm{x}_2) = - \sum_{i = 1}^{M_2} \pdv{}{x_2^i} \left[\sum_{j=1}^{M_2}W_{22}^{ij} \phi_{22}(x_2^j - \theta_2^j) - \sum_{k = 1}^{M_1} A_{21}^{ik}\int d\bm{x}_1 \, p^\mathrm{st}_{1 | 2}(\bm{x}_1 \, | \, \bm{x}_2) \phi_{21}\left(x_1^k-\theta_1^k\right) \right] + \sum_{i = 1}^{M_2} D_2^i\pdv[2]{}{{x_{2}^{i}}}
\end{equation}
so that the effect of the fast timescale becomes that of an extra nonlinear term for $\bm{x}_2$. That is, the fast layer relays the nonlinear interactions back to the second layer, generating a self-interaction term.

In the case of linear interactions, we have once more
\begin{equation}
    \mathcal{L}_{1|2}(\bm{x}_2) = -\sum_{i = 1}^{M_1}\pdv{}{x_1^i} \sum_{j=1}^{M_1} W_{11}^{ij}\left[x_1^j - \theta_1^j + \sum_{k = 1}^{M_1}\sum_{l = 1}^{M_2}(W_{11}^{-1})^{jk}A_{12}^{kl}\left(x_2^l - \theta_2^l\right) \right] + \sum_{i =1}^{M_1}D_1^i \pdv[2]{}{{x_{1}^{i}}} 
\end{equation}
where the feedback interactions coming from the second layer act as a drift that explicitly depends on $\bm{x}_2$:
\begin{equation}
    p^\mathrm{st}_{1 | 2}(\bm{x}_1 \, | \, \bm{x}_2) = \mathcal{N}_1\left(\bm{m}_{1|2}(\bm{x}_2), \hat{S}_1\right), \quad \bm{m}_{1|2}(\bm{x}_2) = \bm\theta_1 - \hat{W}_{11}^{-1} \hat{A}_{12} \left( \bm{x}_2 - \bm{\theta}_2 \right)
\end{equation}
with $\hat{W}_{11}\hat{S}_{1} + \hat{S}_{1} \hat{W}_{11}^T = 2 \hat{D_1}$. Then, the effective operator of the slow layer becomes
\begin{align}
    \mathcal{L}_2^\mathrm{eff}(\bm{x}_2) & = - \sum_{i = 1}^{M_2} \pdv{}{x_2^i} \left[\sum_{j=1}^{M_2}W_{22}^{ij} (x_2^j - \theta_2^j) - \sum_{k = 1}^{M_1} A_{21}^{ik}\int d\bm{x}_1 \, p^\mathrm{st}_{1 | 2}(\bm{x}_1 \, | \, \bm{x}_2) \left(x_1^k-\theta_1^k\right) \right] + \sum_{i = 1}^{M_2} D_2^i\pdv[2]{}{{x_{2}^{i}}} \nonumber \\
    & = -\sum_{i = 1}^{M_2} \pdv{}{x_2^i} \left[\sum_{j=1}^{M_2}W_{22}^{ij} (x_2^j - \theta_2^j) - \sum_{k = 1}^{M_1} A_{21}^{ik} \left(m^k_{1|2}(\bm{x}_2)-\theta_1^k\right) \right] + \sum_{i = 1}^{M_2} D_2^i\pdv[2]{}{{x_{2}^{i}}}
\end{align}
where
\begin{equation}
    m^k_{1|2}(\bm{x}_2)-\theta_1^k = -\sum_{i = 1}^{M_1} \sum_{j = 1}^{M_2} (W_{11}^{-1})^{ki}A_{12}^{ij}\left(x_2^j - \theta_2^j\right) \; .
\end{equation}
This term acts as a self-interaction term with an interaction matrix $\hat{W}_\mathrm{loop} = \hat{A}_{21}\hat{W}_{11}^{-1} \hat{A}_{12}$ which explicitly shows that the feedback interactions are passed through the dynamics of the fast layer, described by $\hat{W}_{11}$, and then instantaneously propagated back to the slow layer, as expected. Thus
\begin{equation}
    p^\mathrm{eff}_{2}(\bm{x}_2) = \mathcal{N}_2 \left(\bm{\theta}_2, \hat{S}_2^\mathrm{eff}\right), \quad \left(W_{22}^{ij} + W_\mathrm{loop}^{ij}\right)\hat{S}_2^\mathrm{eff} + \hat{S}_2^\mathrm{eff} \left(W_{22}^{ij} + W_\mathrm{loop}^{ij}\right)^T = 2 \hat{D_1}
\end{equation}
is its stationary solution. 

\begin{figure}
    \centering
    \includegraphics[width=\textwidth]{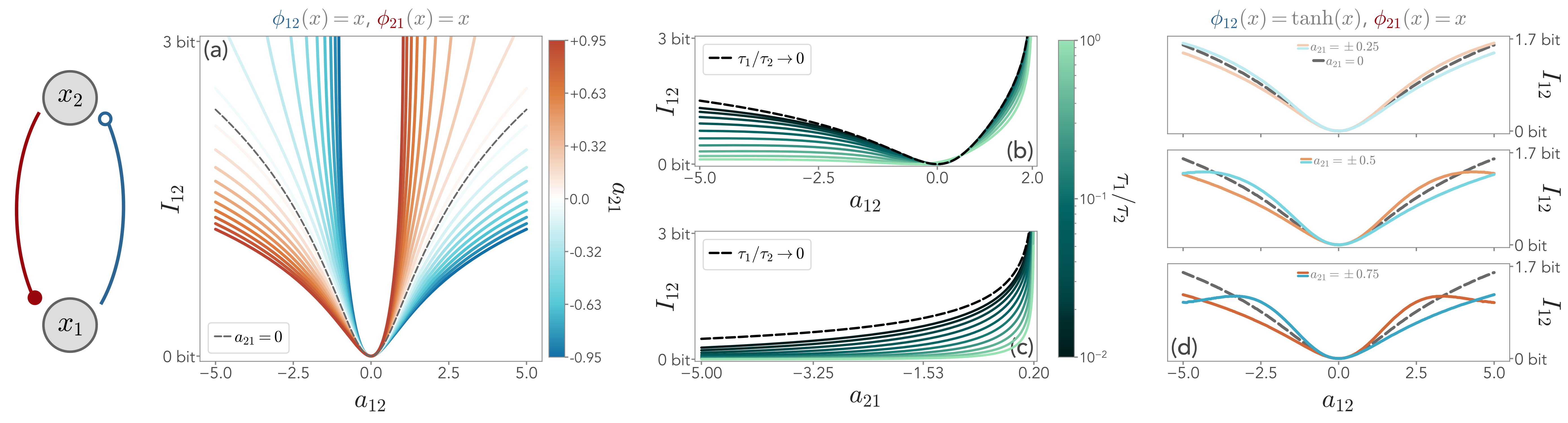}
    \caption{(a) Mutual information $I_{12}$ between a fast and a slow degree of freedom influencing each other, as a function of the coupling from the slow to the fast layer ($a_{12}$). The color-code indicates different values of $a_{21}$, i.e., the fast to slow coupling. For linear interactions, with respect to the baseline where no direct interaction is present ($a_{21} = 0$, gray dashed line), a positive $a_{21}$ is beneficial when $a_{12} > 0$, and detrimental for $a_{12} < 0$. The opposite is true for negative $a_{21}$. (b) Mutual information at fixed $a_{21} = 1/2$ and at different relative temporal scales $\tau_1/\tau_2$. The mutual information tends to be larger in the timescale separation limit $\tau_1/\tau_2 \to 0$ (black dashed line). (c) Same, but at fixed $a_{12} = 5$. (d) Comparison between the linear case (dashed lines) and the nonlinear activation function $\phi_{12}(x) = \tanh(x)$ (solid lines). We find qualitatively similar results. In particular, $I_{12}$ remains well-defined beyond the region of linear stability, and is pushed towards smaller values for large couplings due to the saturating effects of the activation function. In this figure, all diffusion coefficients have been set to $1$ and all biases $\theta_\mu^i = 0$.}
    \label{fig:loop}
\end{figure}

In Figure \ref{fig:loop}a, we plot the results for two one-dimensional layers. In the linear case, we find that the direct interactions is beneficial when the coupling $a_{21}$ has the same sign as $a_{12}$, the feedback. This remains true until the point of linear instability $a_{12}a_{21} = 1$, where the mutual information diverges \cite{barzon2025excitation}. Notably, in Figure \ref{fig:loop}b-c, by solving the system at generic timescales $\tau_1$ and $\tau_2$, we show that $I_{12}$ tends to be larger in the timescale separation limit $\tau_1 / \tau_2 \to 0$.

Finally, we consider the case of nonlinear feedback interactions, where $\phi_{12}(x) = \tanh(x)$. In this case, since $p^\mathrm{st}_{1 | 2}$ remains Gaussian with mean $a_{12}\tanh(x_2)$, the slow layer is described by the effective Langevin equation
\begin{equation}
    \dot{x}_2(t) = - x_2(t) +a_{21} \, a_{12}\tanh(x_2) + \sqrt{2} \,\xi_S(t)
\end{equation}
where we set the diffusion coefficients to $1$ for simplicity. For each $x_2(t)$, the corresponding $x_1(t)$ can be obtained by the change of variable $x_1(t) = a_{12} \tanh(x_S(t)) + \xi(t)$, with $\xi(t) \sim \mathcal{N}(0,1)$. This allows us to sample directly the marginal distribution of the fast layer. Therefore, the mutual information is simply given by $I_{12} = H_1 - 1/2 \log(2\pi e)$, where the differential entropy $H_1$ can be efficiently estimated numerically from the samples $x_1(t)$ \cite{kraskov2004estimating,nicoletti2024integration}. In Figure \ref{fig:loop}d, we show that $I_{12}$ is now well-defined beyond the region of linear stability. In particular, while for small couplings the sign of $a_{12}$ and $a_{21}$ play the same role as in the linear case, for larger values the mutual information diminishes due to the saturating effects of the activation function.

\section{Connection between continuous- and discrete-state systems}
\noindent Let us remark once again that the derived decomposition of the probability distribution is independent of the underlying dynamics and solely relies on the assumption of a sufficient separation between the different timescales at play. Moreover, our result for stochastic differential equations mirrors the one obtained for discrete-state systems with higher-order interactions between nodes and edges, i.e., triadic interactions \cite{sun2023dynamic,niedostatek2024mining} (see \cite{nicoletti2024information} for the derivation of the timescale separation approach in this case). Here, we show that the presence of these higher-order couplings in the discrete-state scenarios can be mapped into a set of coupled Langevin dynamics, as for Eq.~\eqref{eqn:langevin_clusters}. In particular, this parallel equally holds even for linear Langevin equations, a particularly simple setting that can be explicitly solved without timescale separation and has been extensively studied in \cite{nicoletti2024gaussian}.

To start, let us recall the exact definition of triadic interactions between different layers in a discrete-state system, following. Crucially, each layer must represent now distinct degrees of freedom, so that we can identify the state of a layer with the occupied node. To this end, consider $N$ random walkers, one for each layer $\mu$, whose state is $x_\mu^i$, with $i$ a node belonging to layer $\mu$. The dynamics of the random walker in layer $\mu$ is governed by a transition matrix $\hat{W}_\mu$, which plays the same role as $\mathcal{L}_\mu$. Interactions between different layers can be described as dependencies of $\hat{W}_\mu$ on the states of random walkers in other layers, i.e., $\hat{W}_\mu(\bm{x}_1,\dots,\bm{x}_N)$. Since the layers now represent the state space of physically separated systems, interactions between them cannot be pairwise to maintain the distinguishability of states. Rather, the dependencies of $\hat{W}_\mu$ should be interpreted as interactions stemming from the nodes of other layers (the states) to the edges of layer $\mu$ (the transition rates). These higher-order interactions are known as triadic interactions. A mathematically rigorous discussion of this setting can be found in \cite{nicoletti2024information}. To make the calculations simpler, and without loss of generality in the idea, here we consider the situation in which each layer, decoupled from the others, is constituted by a ring governed by a Master Equation dynamics. To simplify the notation even further, consider the following rates:
\begin{equation}
    w_{\mu}^{i\to i+1} = \left( 1 + \frac{\alpha_\mu}{2} \Delta x \right) \frac{w_\mu}{\Delta x^2} \qquad w_{\mu}^{i+1\to i} = \left( 1 - \frac{\alpha_\mu}{2} \Delta x \right) \frac{w_\mu}{\Delta x^2} \;.
\end{equation}
where $\Delta x$ encodes the discretization of the state-space. By performing a continuum limit ($\Delta x\to 0$) in each layer \cite{busiello2019entropy}, we end up with $N$ decoupled Langevin dynamics with drift $A_\mu = \alpha_\mu w_\mu$ and diffusion $D_\mu = w_\mu$, satisfying periodic boundary conditions. Introducing node-dependent rates will result into a drift and diffusion coefficient that depend on $\bm{x}_\mu$. Similarly, introducing arbitrary triadic interactions between layers, i.e., $w_\mu^{i\to j} \to w_\mu^{i\to j}(\bm{x_1},\dots,\bm{x}_N)$, will lead to a dependence of drift and diffusion on the states of other layers, $\bm{x}_\nu$ with $\nu \neq \mu$.

An interesting case is represented by pairwise-connected Langevin equations. Indeed, while triadic interactions can be thought of as higher-order couplings, these can be considered on the same footing as pairwise couplings in continuous systems, at least for the sake of the presented framework. To illustrate this, consider the following choice for the transition rates \cite{busiello2019entropy}:
\begin{equation}
    w_\mu^{i\to i\pm1} = \left( 1 \pm \frac{\sum_{i=1}^N \sum_{j=1}^{M_j} \alpha_\mu^{ij} x_i^j}{2} \Delta x \right) \frac{w_\mu}{\Delta x^2}
\end{equation}
which leads to
\begin{equation}
    A_\mu = \sum_{i=1}^N \sum_{j=1}^{M_j} \alpha_\mu^{ij} x_i^j \qquad D_\mu = w_\mu
\end{equation}
which is exactly the case of Gaussian processes with pairwise couplings across different timescales studied here and in \cite{nicoletti2024gaussian}. 
As such, the discrete-state scenario constitutes, to some extent, a generalization of the continuous case, compatibly with the fact that coarse-graining procedures typically cause loss of information in the system \cite{busiello2019entropy,cocconi2022scaling,yu2021inverse}.

\section{Discussion}
\noindent In this work, we presented a multiple-timescale separation approach to study complex multiscale stochastic systems. This procedure shows that the structure of interactions across timescales determines the conditional structure of the probability distribution, independently of the specific underlying dynamics. This result has important consequences for the mutual information across timescales. In particular, we were able to highlight three basic principles governing information propagation. These, in turn, reveal that feedback interactions, i.e., those from slow to fast layers (e.g., regulatory mechanisms), are able to generate information. On the contrary, direct interactions, i.e., those from fast to slow degrees of freedom, cannot generate information but only propagate it to other layers. Generalizing previous results to the case of nonlinear Fokker-Planck operators, this work paves the way for the study of several information-processing systems and multiscale stochastic processes in general. For instance, ecological processes are shaped by the interplay between the timescale of population dynamics and environmental changes \cite{hastings2018transient, francis2021management, nicoletti2023emergence, arnoulx2024many, pichon2024integrating} or exploration and settlement \cite{nicoletti2023emergent, doimo2025finite}, while a wide range of timescales define neural dynamics and are believed to support function and computation \cite{cavanagh2020diversity, gao2020neuronal, zeraati2023intrinsic, zeraati2024neural, senkowski2024multi}. Also, reaction-diffusion systems exhibit emergent features that are fundamentally shaped by the interplay between processes acting at different timescales, such as the spatial selection of target species in chemical \cite{agudo2020cooperatively,astumian2014enhanced,mandal2023kinetic} or thermal gradients \cite{piazza2008thermophoresis,liang2022emergent,duhr2006optothermal,braun2002trapping}, and the onset of complex patterns driven by chemical activity \cite{liang2024thermodynamic,wurthner2022bridging,denk2018mine,brauns2020phase}. For these reasons, we believe that the role of the underlying multiscale structure will prove fundamental in studying and determining the performance of encoding and propagating information in complex systems.

\begin{acknowledgments}
\noindent The content of this work has been inspired by the authors' presentations at the 29th International Conference on Statistical Physics (STATPHYS29) in Florence, Italy, and is an invited submission to the conference proceedings.
\end{acknowledgments}


%

\end{document}